# Artifacts in magnetic spirals retrieved by transport of intensity equation (TIE)

J. Cui[a, b], Y. Yao[a, *], X. Shen[a], Y. G. Wang[a], R. C. Yu[a, b]

a. Beijing National Laboratory of Condensed Matter Physics,

Institute of Physics, Chinese Academy of Sciences,

Beijing 100190, China

b. School of Physical Sciences, University of Chinese Academy of Science

Beijing 100049, China

*corresponding author: yaoyuan@iphy.ac.cn

**Abstract:**

The artifacts in the magnetic structures reconstructed from Lorentz transmission electron microscopy (LTEM) images with TIE method have been analyzed in detail. The processing for the simulated images of Bloch and Neel spirals indicated that the improper parameters in TIE may overestimate the high frequency information and induce some false features in the retrieved images. The specimen tilting will further complicate the analysis of the images because the LTEM image contrast is not the result of the magnetization distribution within the specimen but the integral projection pattern of the magnetic induction filling the entire space including the specimen.

## 1. Introduction

Lorentz transmission electron microscopy (LTEM) has been known as an important characterization method to reveal micro magnetic structures in magnetic materials and devices. At the Fresnel mode, the white or black stripes in LTEM images demonstrate the magnetic features within specimens [1]. However, those stripes could not reveal the details of magnetic structures directly because LTEM images are acquired at a large defocus and it is difficult to recognize the relationship among magnetic patterns only from the contrast. Transport of intensity equation (TIE) has been applied to deal with the LTEM images to disclose the magnetic induction and then infer the magnetization direction [2, 3]. Compared with electron holography (EH) which can also retrieve the magnetic information of the specimens in transmission electron microscopy (TEM), the advantages of TIE are the simpler experiment framework and faster data processing. Now TIE has achieved a widespread success in portraying domains within magnetic materials and devices [4-8]. The discovery of the so-called Skyrmions in some materials demonstrates the visualization advantage of TIE. The subsequent literature has reported various spirals with the assistance of TIE [9-15]. Nagaosa et al [16] classified those spiral species according to their chirality. However the possible artifacts in TIE are seldom mentioned in the published literature, especially for magnetic structures less than hundreds of nanometers. Here the artifacts of TIE were analyzed to illustrate the potential misleading due to the incaution in data processing.

## 2. The principle of TIE

TIE is an approach to recover the phase of the wave exiting from the specimen. Teague [17] and Paganin et al [18] established this method by analyzing the general relationship between intensity $I(\vec{r}_\perp, z)$ and phase $\varphi(\vec{r}_\perp, z)$ of a wave which propagates in the free space,

$$\varphi(\vec{r}_\perp, z_0) = -k\nabla_\perp^{-2}\left\{\nabla_\perp\left[\frac{\nabla_\perp \nabla_\perp^{-2}\left(\frac{\partial I(\vec{r}_\perp, z)}{\partial z}\right)}{I(\vec{r}_\perp, z_0)}\right]\right\} \quad (1)$$

where $\vec{r}_\perp$ is the coordinate in the plane perpendicular to the propagation direction $z$ and $k$ is the wave vector. In this way, the phase can be retrieved directly from the variation of $I(\vec{r}_\perp, z)$ at different positions along the transmission direction. The Fourier transfer (FT) method can be used to solve the partial differential equation (eq. 1) by replacing the inverse Laplacian $\nabla_\perp^{-2}$ [19]

$$\nabla_\perp^{-2} f(x,y) = \mathcal{F}^{-1}\left\{\frac{\mathcal{F}[f(x,y)]}{|q_\perp|^2}\right\} \quad (2)$$

where $q_\perp$ is the spatial frequency in the plane which is perpendicular to the beam direction. However, eq. 2 is divergent and the noise is magnified when $q_\perp$ is nearby zero, thus it should be modified by adding a small nonzero constant $q_0$ to avoid the divergence and suppress low-frequency noises [19, 20]. But the cost is obvious – both the magnitude and frequency dependence could be altered inevitably by $q_0$.

$$\nabla_\perp^{-2} f'(x,y) = \mathcal{F}^{-1}\left\{\frac{\mathcal{F}[f(x,y)]}{|q_\perp|^2 + q_0^2}\right\} \quad (3)$$

The flow chart to solve eq. 1 with FT method is shown in fig. 1. It should be emphasized that eq. 3 is applied twice during the phase reconstruction. Because of its convenience, TIE has been utilized successfully in many TEM researches, particularly in LTEM which characterizes magnetic microstructures by obtaining images at hundreds of micrometers defocus [4, 21].

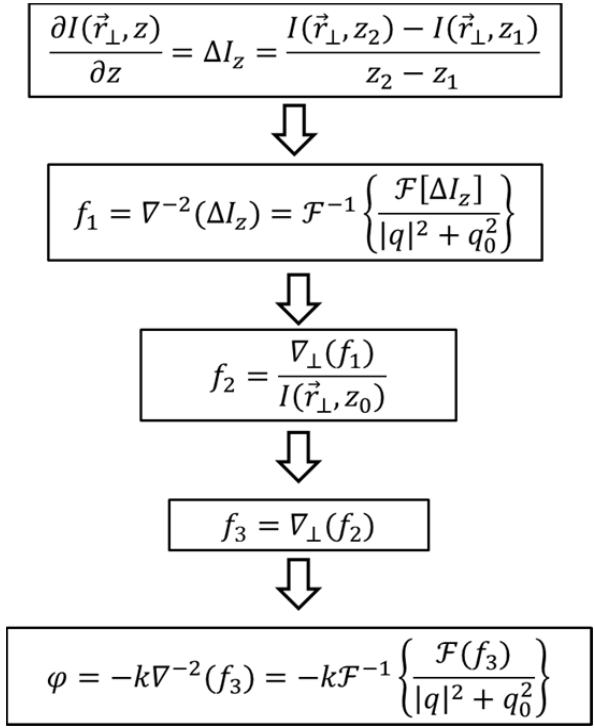

**Fig. 1.** The flow chart of the FT aided TIE algorithm.

## 3. The simulation of magnetic spirals and LTEM images

Finite element method (FEM) software COMSOL was employed to simulate the magnetic induction of Bloch and Neel type spirals in the materials with the given magnetization arrangement. The geometric figure of the spirals is shown in fig. 2a and more details are described in Appendix. The magnetization was $M_0$ in the domain core but $-M_0$ outside the domain wall. The magnetization $\boldsymbol{M}$ altered its orientation $\theta$ through the domain wall in the form given by eq. A3, as shown in fig. 2b and 2c. COMSOL can calculate the final distribution of the magnetic field $\boldsymbol{B}$ in the entire space including the specimen. Fig. 2b and 2c illustrate the calculated magnetic induction of the Bloch type and Neel type spirals, respectively. The specimen simulated in COMSOL was rotated on $y$ axis perpendicular to the electron beam to mimic the sample tilting in LTEM.

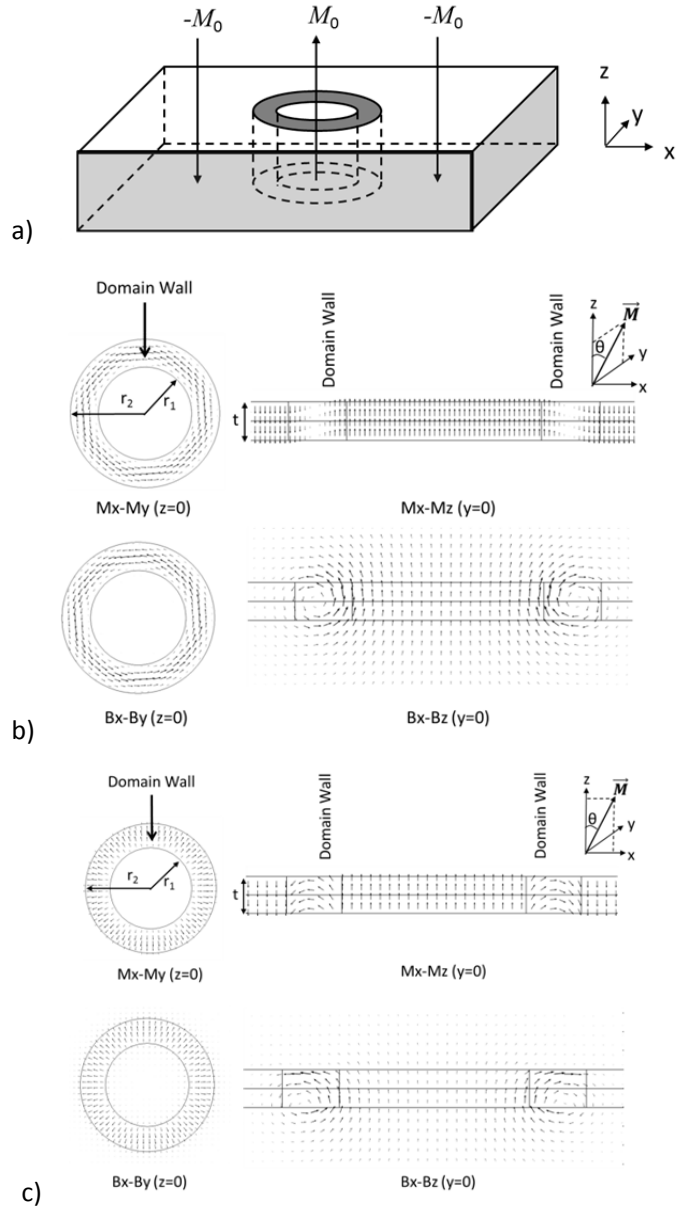

**Fig. 2.** a) The schematic of the domains with an annular wall. The electron beam moved along *z* direction and the specimen was titled on *y* axis. b) and c) The magnetization arrangement and the calculated magnetic induction of the Bloch type and Neel type, respectively. (Left: *x-y* plane, right: *x-z* plane.)

If the contribution from the electrostatic potential could be ignored, the contrast of the LTEM image is determined mainly by the convolution of contrast transfer function (CTF) [22] of the objective lens and the exit electron phase which relies on the integral of *z* component of the vector potential $A_z(\vec{r}_\perp, z)$ (eq. 4) [23]. However, a large defocus (more than several hundred micrometers) is needed to form the discernible contrast because the phase shift induced by the magnetic field is too small. This imaging is known as the Fresnel mode. A long-focal-length objective lens that does not influence the magnetic sample is used to image the exit wave in Lorentz microscope, but unfortunately, its spherical aberration Cs is about several meters, three orders of the magnitude for the conventional objective lens in the ordinary high resolution TEM (HRTEM). So the spatial resolution of the LTEM is not better than a few nanometers [24].

$$\varphi(\vec{r}_\perp) = -\frac{e}{\hbar} \int_{-\infty}^{+\infty} A_z(\vec{r}_\perp, z) \mathrm{d}z \qquad (4)$$

eq. 4 also indicates another important fact that the contrast in the image depends on the in-plane component of the magnetic field, not the magnetization, though a so-called "projected" magnetization $M'(x, y)$ can be deduced from the recovered phase $\varphi$ [25]

$$M'_x \propto B'_x \propto -\frac{\partial \varphi}{\partial y}, \ M'_y \propto B'_y \propto -\frac{\partial \varphi}{\partial x} \qquad (5)$$

If only the relative strength and direction of magnetization are concerned, it is thought that the partial difference of $\varphi$ (eq. 5) may be used to directly estimate the projected $M'(x, y)$, which is widely applied in many literature.

The integral of $A_z(\vec{r}_\perp, z)$ (eq. 4) was implemented easily with COMSOL built-in operator to generate a two-dimensional phase map，then that phase map was input into a homemade Digitalmicrograph (DM) script to simulate the TEM images with different optical parameters. TIE script recovered the phase images again from the simulated images to investigate the influence of the parameter $q_0$ in the TIE processing.

Some typical simulated LTEM images with different tilting angles and defocuses are displayed in Fig. 3. There is no contrast in the images for Neel type spirals without tilting because the magnetic fluxes are opposite on the top and bottom surfaces of the domain wall (fig. 2c) and the total phase change of the transmission electron wave was almost zero. When the specimen was tilted, this symmetry was broken (Fig. 10) and the fluxes could not cancel each other in the *x-y* plane so that the contrast emerged.

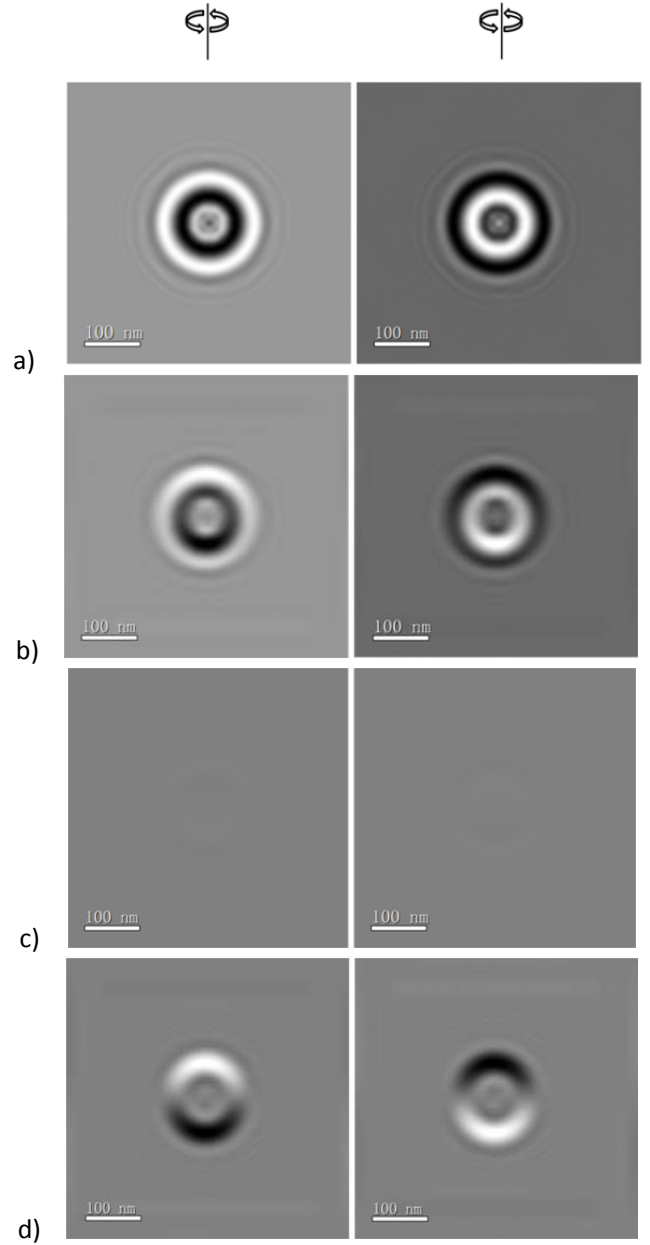

**Fig. 3.** The simulated LTEM images for the spiral structures at different defocuses. a) and b) Bloch type without tilting and with 20° tilting; c) and d) Neel type without tilting and with 20° tilting. (Left: defocus -500 μm, right: defocus 500 μm, accelerate voltage: 200 keV, Cs: 5 m, Cc: 100 Å)

## 4. Retrieve of the projected magnetization direction

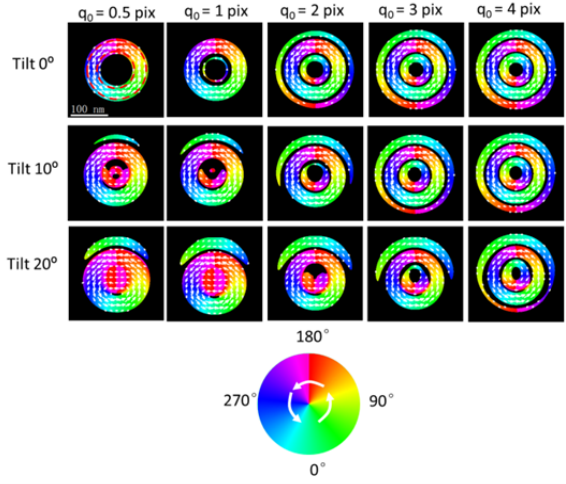

**Fig. 4.** TIE retrieved magnetization $M'(x,y)$ of a Bloch type spiral for different $q_0$ and specimen tilting. The red dash circles indicate the boundary of domain walls. (1 pix = 1/512 nm$^{-1}$ ≈ 0.002 nm$^{-1}$)

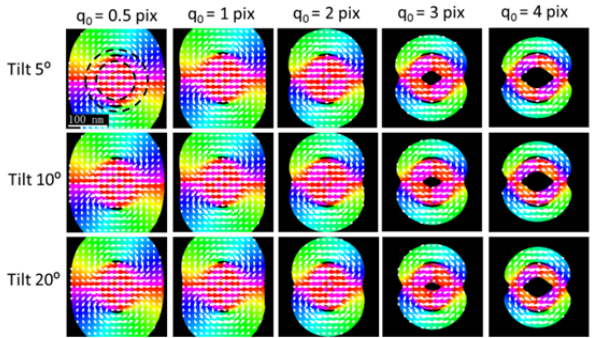

**Fig. 5.** TIE retrieved magnetization $M'(x,y)$ of a Neel type spiral for different $q_0$ and specimen tilting The black dash circles indicate the boundary of domain walls. The patters of 5° tilting substitute those of 0° tilting because the later show no contrast in the simulation images. (1 pix = 1/512 nm$^{-1}$ ≈ 0.002 nm$^{-1}$)

Fig. 4 and 5 illuminate the orientation maps of the projected magnetization $M'(x,y)$ extracted according eq. 5 from the TIE recovered phases of Bloch and Neel spirals, respectively. In order to reveal the influence of $q_0$, different $q_0$ was used to retrieve the phase patterns of the tilted specimen. It is apparent that $q_0$ can modify the resulted $M'(x,y)$ patterns. For the Bloch spiral without tilting, the fidelity is good when $q_0$ is small – the recovered magnetizations locating within the domain wall is as same as the setting state. But as $q_0$ increasing, a corona with the inverse magnetization first emerges inside the domain wall and the gradually spreads to the outside where there should not be any in-plane magnetic magnetization components. Moreover, the specimen tilting leads to an image distortion and turns the corona into the arches. For the Neel spiral, a small tilting can result in the double spiral with inverse chirality. Those double spirals do not exist in the magnetization setting of the Neel type model in fig. 2c. Changing $q_0$ also resulted in the features variation. Therefore, the improper $q_0$ and specimen tilting complicate the analysis of the magnetization patterns.

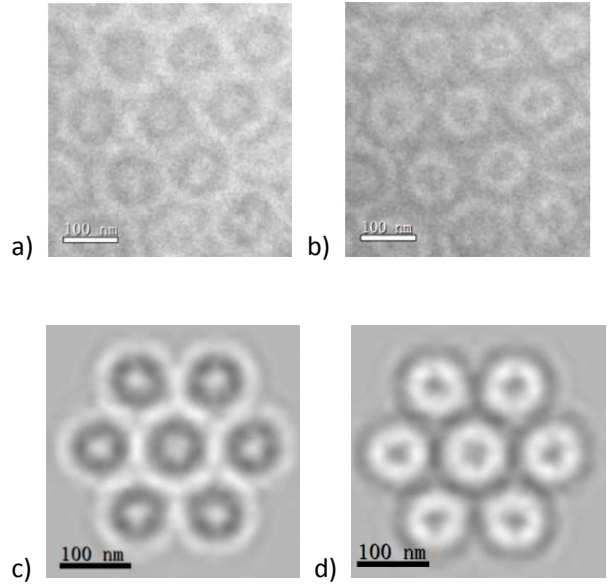

**Fig. 6.** a), b) LTEM images of the magnetic spirals in NiMnGa foil (defocus: -300 μm and 300 μm, respectively), c) ,d) the simulation images of an array with six Bloch spirals (defocus -500 μm and 500μm, respectively).

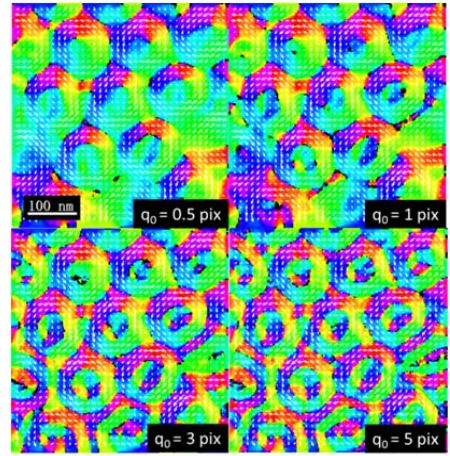

**Fig. 7.** The projected magnetization $M'(x,y)$ retrieved from the experimental images in fig. 6 a) and b) with different $q_0$.

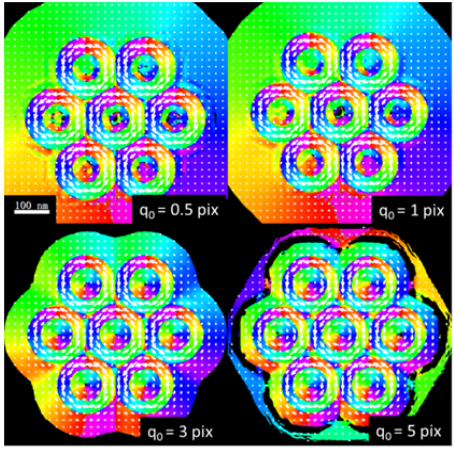

Fig. 8. The projected magnetization $M'(x,y)$ retrieved from the simulation images in fig. 6 c) and d) with different $q_0$.

A NiMnGa alloy was characterized at Fresnel mode in Tecnai F20 equipped with Lorentz lens to investigate the artifacts in TIE processing for experimental data. Fig. 6 shows the images acquired at different defocuses, together with the simulations of a Bloch spiral array (the physical structure of the array is depicted in the Appendix). The retrieved projected magnetization $M'(x,y)$ with different $q_0$ (fig. 7 and fig. 8) exhibit the similar features as those shown in fig. 4, implying the Bloch type spiral array exists in the specimen.

5. Discussion

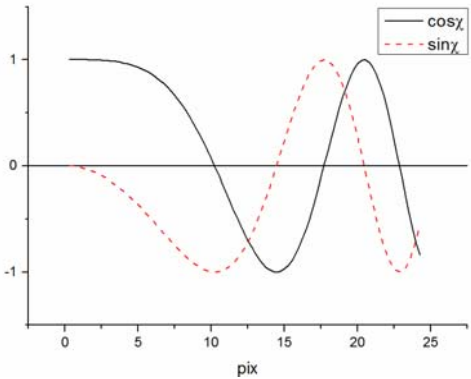

Fig. 9 CTF curves for the LTEM image simulation parameters. (Accelerate voltage: 200 keV, Cs: 5 m, defocus: 500 nm)

Eq. 3 denotes that the Fourier method of TIE solution should distort the recovered phase pattern due to $q_0$. Using $q_0$ is equivalent to introducing a high-pass filter to suppress the low frequency noise and avoid the divergence at zero point [20] but the low frequency signal may be lost and the high frequency part should be magnified improperly, although larger $q_0$ can highlight the image detail since high frequency signal carries the fine structures in an image. Moreover, the projected $M'(x,y)$ is sensitive to the spatial frequency as it is calculated from the partial difference of the phase in which rapid spatial change implies a strong magnetic field (eq. 5). So when the low frequency noise is damped by the filter in the phase recovery, the exaggerated high frequency information may enlarge the original weak magnetic signals and distorts the magnetization mapping, as the artifacts displayed in fig. 4 and fig. 5. Meanwhile, the image contrast is the convolution of the exit wave and CTF which is also a frequency related function (Fig.9), $q_0$ may enhance the high frequency component in CTF and deteriorate the reconstructed results too.

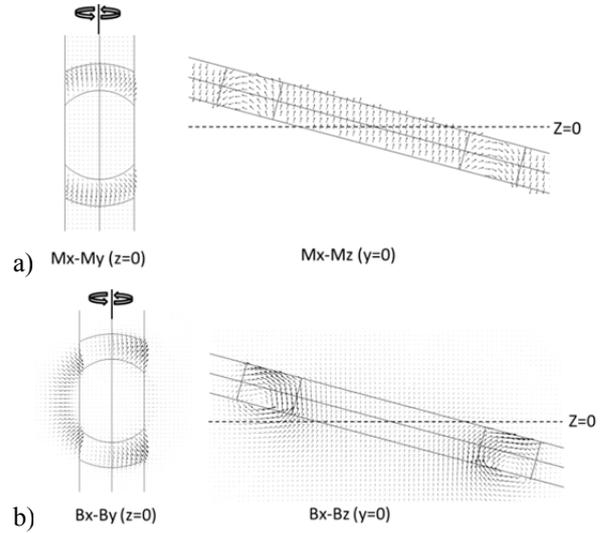

Fig. 10 The difference between a) magnetization **M** and b) magnetic induction **B** for a 15° tilting Neel type spiral. The solid lines outline the boundaries of the domains and the walls in the corresponding cross section.

Another misguidance is from that the contrast of the images acquired in LTEM is determined by the integral projection of the magnetic field. The phase of the electron beam relates with the magnetic induction **B** in the space, not the magnetization **M**. The retrieved magnetic information cannot directly reflect the real three dimensional structures, especially when the specimen is tilted, as indicated in fig. 10 where the difference between **B** and **M** is demonstrated too. Magnetic fluxes may exist in the space even there are not any magnetizations. When the sample is tilted the magnetic fluxes which are initially along the electron beam can produce the horizontal component and the unsymmetrical flux distribution brings the phase contrast in the images (Fig. 3d). From the $B_x - B_y$ diagram in Fig. 10b it is easy to find a double spiral which is generated by the combination of the stray field in the free space and the magnetic field in the specimen. Thus the double-spirals in the projected $M'(x,y)$ mappings of the tilted Neel structure are the exhibitions of the integral projection of the magnetic induction $\mathbf{B}$, not the actual distribution of the magnetization $\mathbf{M}$ in the specimen. Considering that the magnetic induction fills the space involving the specimen and the stray field cannot be avoided or got rid of in LTEM at all, eq. 5 should be utilized with caution when deducing the magnetization in the specimen.

The double spiral originated from the magnetic induction distribution is independent of the magnetization rolling detail in the domain walls to some extent because the inductions should be closed regardless of what the magnetic configuration is in the above geometric figures. The double spiral generated from a different magnetization rotation mode also proves this robustness (fig. A4). Pollard et al has reported the similar phenomenon for the Neel type Skyrmion recently [26].

## 6．Conclusion

TIE is a convenient approach in LTEM to uncover the tiny magnetic structures due to its simple data processing and visualization. But the caution should be paid during the application because the parameters in the image processing may lead to some artifacts. Improper large $q_0$ will overestimate the contribution from high frequency signals and result in the misleading magnetic structures in the output. Some no-existent magnetizations gradually appear in the retrieved patterns as $q_0$ increasing. The nature that the image in LTEM only responds to the integral of the in-plane magnetic induction also hinders the accurate analysis of the real magnetization in the specimen. A simple magnetic structure may demonstrate various features when it changes its orientation in LTEM. This complicates the understanding of the experimental data, especially when the prior knowledge and the orientation of the magnetic configuration in LTEM are unknown.

## Appendix

1. Geometry of the simulated specimen

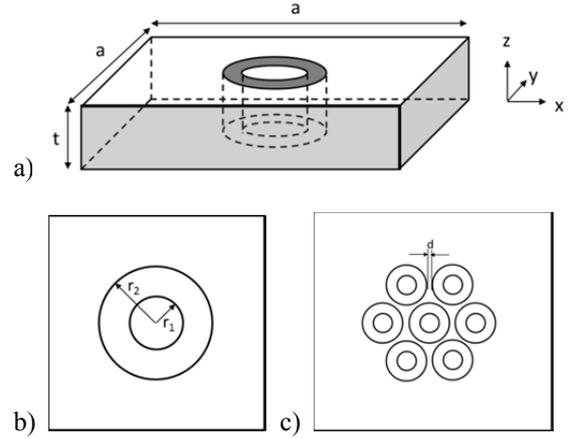

Fig. A1 The configurations of the specimen simulated in COMSOL.

Fig. A1 displays the geometric structures of the specimen for the simulation. $r_1$ was 50 nm and $r_2$ was 80 nm for the single Bloch or Neel spirals, respectively. The spirals located inside a block with size $a \times a \times t$ (Fig. A1a). $a$ was 400 nm but $t$ was 100 nm (Bloch type) or 20 nm (Neel type) (Fig.A1b). For the Bloch spiral array, $r_1$ was 30 nm and $r_2$ was 50 nm, respectively, with the enlarged $a$ 800 nm, $t$ 100 nm and the spacing $d$ 10 nm between the spirals (Fig.A1c). The block was surrounded by a large air cubic. Four side surfaces of the block (gray surfaces in Fig.A1a) were defined as magnetic shielding to avoid the edge effect. The relative magnetic permittivity $\mu_r$ was 1000.

2. Magnetic configuration

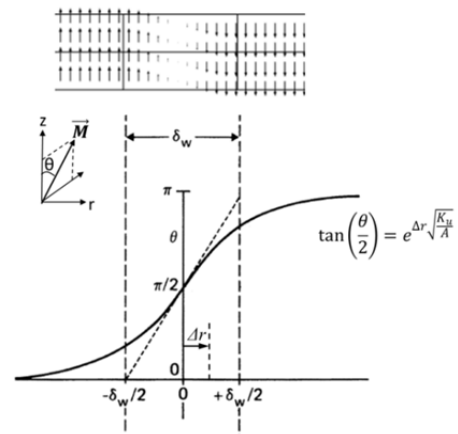

Fig. A2 The configuration of the magnetization in the domain wall.

The magnetic configurations in the domain wall of the Bloch or Neel type spirals followed the description in ref. [27]. The rotation $\theta$ of the magnetization $\mathbf{M}$ in the domain wall is

$$\tan\left(\frac{\theta}{2}\right) = e^{\Delta r \sqrt{\frac{K_u}{A}}}, \tag{A1}$$

where $\Delta r$ is the distance from the center of the domain wall, $K_u$ and $A$ are the unaxial anisotropy parameter and exchange constant, respectively. The thickness $\delta w$ of the wall is defined as

$$\delta w = \pi \sqrt{\frac{A}{K_u}}. \tag{A2}$$

Thus

$$\tan\left(\frac{\theta}{2}\right) = e^{\pi \frac{\Delta r}{\delta w}}. \tag{A3}$$

It means that the rotation of magnetization in the wall can be calculated from its position and the wall thickness $\delta w$ directly. For cobalt, $K_u$ is $5.2 \times 10^5$ J/m$^3$ and $A$ is $3 \times 10^{-11}$ J/m, so $\delta w$ is about 23 nm. For convenience, $\delta w$ was set to 30 nm for the single spiral and 20 nm for the spiral array, respectively, to calculate the magnetic field.

3. The influence of the magnetization setting

A different magnetization rotation mode was set to a linearly change from 0 to $\pi$ through the domain wall in single Neel type spiral.

$$\theta = -\frac{\pi}{2} + \frac{\pi}{\delta w}\Delta r \tag{A4}$$

The calculated magnetic inductions are shown in fig. A3. As same as the previous process, the projected $M'(x,y)$ mappings are displayed in Fig. A4. These results are similar to those in fig. 5.

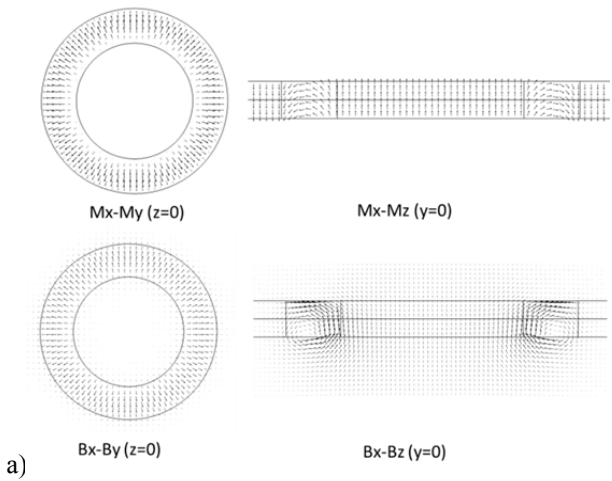

a)

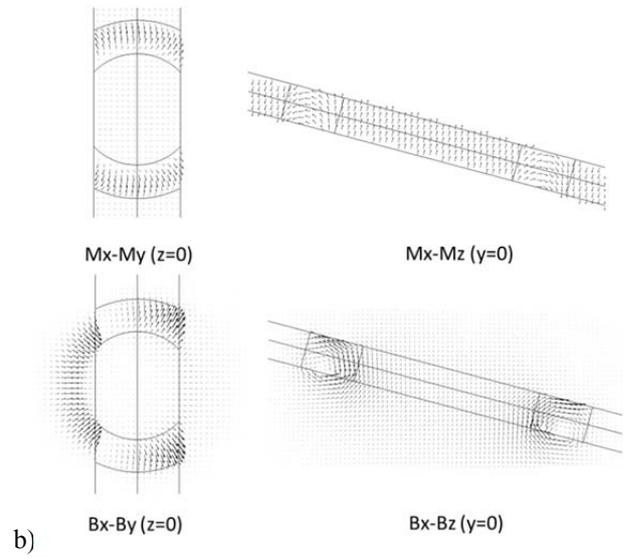

b)

Fig. A3 The magnetization and calculated magnetic induction of the Neel type spiral with configuration eq. A4. a) Without tilting and b) 15° tilting, respectively.

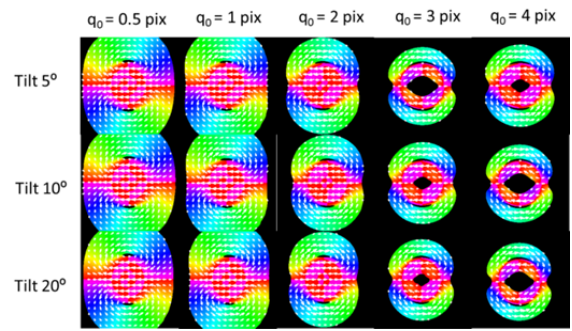

Fig. A4 The projected $M'(x,y)$ retrieved from the simulation images of the Neel type spiral with configuration eq. A4.

**Acknowledgments**

This work was supported by the Ministry of Science and Technology of the People's Republic of China (Grant No. 2016YFA0202500, 2016YFB0300500) and key science project of Jiangsu province, China (Grant No. BE2015216).

**References**

[1] M. S. Cohen, Wave‐Optical Aspects of Lorentz Microscopy, J. Appl. Phys. 38 (1967) 4966-4976

[2] V. V. Volkov, Y. Zhu, Lorentz phase microscopy of magnetic materials, Ultramicroscopy 98 (2004) 271–281

[3] M. Beleggia, M. A. Schofield, V. V. Volkov, Y. Zhu, On the transport of intensity technique for phase retrieval, Ultramicroscopy 102 (2004) 37–49

[4] V. V. Volkov and Y. Zhu, Phase Imaging and Nanoscale Currents in Phase Objects Imaged with Fast Electrons, Phys. Rev. Lett. 91 (2003) 043904

[5] S. D. Pollard, V. Volkov, and Y. Zhu, Propagation of magnetic charge monopoles and Dirac flux strings in an artificial spin-ice lattice, Phys. Rev. B. 85 (2012) 180402(R)

[6] J. Q. He, V. V. Volkov, M. Beleggia, et al, Ferromagnetic domain structures and spin configurations measured in doped manganite, Phys. Rev. B. 81 (2010) 094427

[7] L. A. Rodríguez, C. Magén, E. Snoeck, et al, Optimized cobalt nanowires for domain wall manipulation imaged by in situ Lorentz microscopy, Appl. Phys. Lett. 102 (2013) 022418

[8] E. Humphrey, C. Phatak, A. K. Petford-Long, M. De Graef, Separation of electrostatic and magnetic phase shifts using a modified transport-of-intensity equation, Ultramicroscopy 139 (2014) 5-12

[9] X. Z. Yu, Y. Onose, N. Kanazawa, et al, Real-space observation of a two-dimensional skyrmion crystal, Nature 465 (2010) 901-904

[10] X. Yu, M. Mostovoy, Y. Tokunaga, et al, Magnetic stripes and skyrmions with helicity reversals, Proc. Nat. Acad. Sci. 109 (2012) 8856-8860

[11] S. Seki, X. Z. Yu, S. Ishiwata, Y. Tokura Observation of Skyrmions in a Multiferroic Material, Science 336 (2012) 198-201

[12] D. Morikawa, X. Z. Yu, Y. Kaneko, et al, Lorentz transmission electron microscopy on nanometric magnetic bubbles and skyrmions in bilayered manganites $La_{1.2}Sr_{1.8}(Mn_{1-y}Ru_y)_2O_7$ with controlled magnetic anisotropy Appl. Phys. Lett. 107 (2015) 212401

[13] Haifeng Du, Renchao Che, Lingyao Kong, et al, Edge-mediated skyrmion chain and its collective dynamics in a confined geometry, Nat. Comm. 6 (2015) 8504

[14] C. Phatak, O. Heinonen, M. De Graef, and A. Petford-Long, Nanoscale Skyrmions in a Nonchiral Metallic Multiferroic: $Ni_2MnGa$, Nano Lett. 16 (2016) 1530-6984

[15] W. Wang, Y. Zhang, G. Xu, et al, A Centrosymmetric Hexagonal Magnet with Superstable Biskyrmion Magnetic Nanodomains in a Wide Temperature Range of 100–340 K, Adv. Mater. 28 (2016) 6887–6893

[16] N. Nagaosa and Y. Tokura, Topological properties and dynamics of magnetic skyrmions, Nat. Nanotech. 8 (2013) 899-911

[17] M. Teague, Deterministic phase retrieval: a Green's function solution, J. Opt. Soc. Am. 73 (1983) 1434-1441

[18] D. Paganin, K. A. Nugent, Noninterferometric Phase Imaging with Partially Coherent Light, Phys. Rev. Lett. 80 (1998) 2586-2589

[19] K. Ishizuka and B. Allman, Phase measurement of atomic resolution image using transport of intensity equation, J. Electron Microsc. 54 (2005) 191–197

[20] M. Mitome, K. Ishizuka and Y. Bando, Quantitativeness of phase measurement by transport of intensity equation, J. Electron Microsc. 59 (2010) 33–41

[21] V. V. Volkov, Y. Zhu, M. De Graef, A new symmetrized solution for phase retrieval using the transport of intensity equation, Micron 33 (2002) 411-416

[22] E. J. Kirkland, Advanced Computing in Electron Microscopy (2nd Ed.), Springer, 2010

[23] J. E. Bonevich, G. Pozzi and A. Tonomura, Electron holography of electromagnetic field, in Introduction to Electron Holography, 1st. ed. E. Volkl, L. F. Allard and D. C. Joy, Kluwer Academic/Plenum Publishers, 1999

[24] C. Phatak, A. K. Petford-Long, M. De Graef, Recent advances in Lorentz microscopy, Ultramicroscopy, Curr. Opin. Solid State Mat. Sci. 20 (2016) 107–114

[25] R. E. Dunin-Borkowski, M. R. McCartney, M. Pósfai, et al, Off-axis electron holography of magnetotactic bacteria: magnetic microstructure of strains MV-1 and MS-1, Eur. J. Mineral. 13 (2001) 671–684

[26] S. D. Pollard, J. A. Garlow, J. Yu, et al, Observation of stable Néel skyrmions in cobalt/palladium multilayers with Lorentz transmission electron microscopy, Nat. Commun. 8 (2017) 14761

[27] A. H. Eschenfelder, Magnetic Bubble Technology (2nd Ed.), Springer, 1981